# UFig v1: The ultra-fast image generator

**Silvan Fischbacher [1*], Beatrice Moser [1*], Tomasz Kacprzak [1,2*], Luca Tortorelli [1,3*], Joerg Herbel [1*], Claudio Bruderer [1*], Uwe Schmitt [1,4], Alexandre Refregier [1], Joel Berge [1,5], Lukas Gamper [1], and Adam Amara [1,6]**

**1** ETH Zurich, Institute for Particle Physics and Astrophysics, Wolfgang-Pauli-Strasse 27, 8093 Zurich, Switzerland **2** Swiss Data Science Center, Paul Scherrer Institute, Forschungsstrasse 111, 5232 Villigen, Switzerland **3** University Observatory, Faculty of Physics, Ludwig-Maximilian-Universität München, Scheinerstrasse 1, 81679 Munich, Germany **4** ETH Zurich, Scientific IT Services, Binzmühlestrasse 130, 8092 Zürich, Switzerland **5** DPHY, ONERA, Université Paris Saclay, F-92322 Châtillon, France **6** School of Mathematics and Physics, University of Surrey, Guildford, Surrey, GU2 7XH, UK * These authors contributed equally.

## Summary

With the rise of simulation-based inference (SBI) methods (see e.g. Cranmer et al., 2020), simulations need to be fast as well as realistic. `UFig v1` is a public Python package that simulates astronomical images with exceptional speed, taking approximately the same time as source extraction. This makes it particularly well-suited for SBI methods where computational efficiency is crucial. To render an image, `UFig` requires a galaxy catalog, and a description of the point spread function (PSF). It can also add background noise, sample stars using the Besançon model of the Milky Way (Robin et al., 2003), and run `SExtractor` (Bertin & Arnouts, 1996) to extract sources from the rendered image. The extracted sources can be matched to the intrinsic catalog using the method described in Moser et al. (2024), flagged based on `SExtractor` output and survey masks. Emulators can be used to bypass the image simulation and extraction steps (Fischbacher et al., 2025). A first version of `UFig` was presented in Bergé et al. (2013) and the software has since been used and further developed in a variety of forward modelling applications (Bruderer et al., 2016; Chang et al., 2015; Fischbacher et al., 2025; Herbel et al., 2017; Kacprzak et al., 2020; Moser et al., 2024; Tortorelli et al., 2020, 2021).

## Statement of need

`UFig` is a crucial part of the GalSBI framework. GalSBI is a galaxy population model that is used to generate mock galaxy catalogs for all kinds of cosmological applications such as photometric redshift estimation, shear and blending calibration or to forward model selection effects and measure galaxy population properties. This model is constrained by comparing simulated data to observed data. To accurately compare the simulations with the data, the simulations need to be as realistic as possible. We therefore need to include instrumental and observational effects such as the PSF and the background noise of the data, as well as the survey masks. This can be done by rendering images from the intrinsic GalSBI catalogs and extracting the sources from the images with the same method as for the data.

Since the dimensionality of the parameter space of the galaxy population model is high (around 50 parameters) and the numbers of simulations required to constrain the model is hence large, a fast image generator is crucial to make the inference feasible. `UFig`'s rendering implementation is based on a combination of pixel-based and photon-based rendering methods (see Bergé et al., 2013 for more details). This hybrid approach is one of the key factors behind `UFig`'s fast rendering speed, which is roughly comparable to the time required for source extraction from the images. For the simulations of the Hyper-Suprime-Cam (HSC, Aihara et al., 2018) deep fields presented in Moser et al. (2024) and Fischbacher et al. (2025), the rendering time is between 5 and 10 seconds for a typical image on a single CPU core.

`UFig` is optimized for wide-field galaxy surveys, where saturated bright objects or artifacts are typically masked, and detailed galaxy morphology (e.g., spiral structures) is not critical. As a result, the software includes only a simplified treatment of saturation, does not simulate artifacts such as cosmic rays, and models galaxy light profiles using Sérsic profiles.

At the same time, `UFig` is capable of accurately modelling background noise, including correlated noise from the coaddition process, and the point-spread function (PSF). This is essential for applying `UFig` to galaxy surveys, such as weak lensing, where precise shape measurements are required. This balance makes `UFig` unique in the field of image simulation compared to other software packages, such as `GalSim` (Rowe et al., 2015) and `GalSim`-based packages like `ImSim` (LSST Dark Energy Science Collaboration (DESC), 2024) and the GREAT3 simulations (Mandelbaum et al., 2015), as well as `PhoSim` (Peterson et al., 2020), `Skymaker` (Bertin, 2009), and other GREAT challenge simulations (Bridle et al., 2010; Kitching et al., 2012). To flexibly adapt to different use cases, `UFig` is based on the `ivy` workflow engine and provides plugins for the different steps of the image generation process. The full workflow can then be defined in a single configuration file, where the user can specify which plugins to use and how to configure them, e.g. by defining the PSF or background model, making the image generation process flexible and easy to use. Examples of configuration files can be found in the Advanced UFig tutorial in the `UFig` documentation.

Compared to the first version of `UFig` presented in Bergé et al. (2013), new features and improvements have been added. In Chang et al. (2015), `UFig` was used to model the transfer function of images of the Dark Energy Survey (DES, Dark Energy Survey Collaboration et al., 2016) from intrinsic galaxy catalogs to measured properties. Bruderer et al. (2016) used `UFig` to render DES-like images for which the PSF modeling and the background noise were adapted to the DES data. Furthermore, to ensure a realistic distribution of the stars in the images, a plugin to sample stars from the Besançon model of the Milky Way (Robin et al., 2003) was added (see also Bruderer, 2018 for a comprehensive overview of the `UFig` features at that time). Herbel et al. (2017) constrained a galaxy population model using `UFig`. This galaxy population model was then used to measure cosmic shear in Kacprzak et al. (2020). This effort required major improvements in the background and PSF modelling. The PSF modelling based on a convolutional neural network (CNN) was first presented in Herbel et al. (2018). Tortorelli et al. (2018) and Tortorelli et al. (2021) adapted `UFig` to render images for narrow-band filters in the context of the Physics of the Accelerating Universe (PAU, Benítez et al., 2009; Martí et al., 2014) Survey. Moser et al. (2024) used `UFig` to simulate deep fields of the HSC which required further adaptions for the PSF modelling and the matching of the extracted sources to the input catalog. Finally, Fischbacher et al. (2025) introduced emulators to bypass the image simulation and extraction steps.

A possible workflow using `UFig` could be the following:

1. Define an intrinsic galaxy catalog. An easy way to do this is to use the GalSBI galaxy population model and its catalog generator (Fischbacher et al., 2024). However, the catalog can also be generated manually without using the GalSBI model.
2. Sample stars from the Besançon model with the `UFig` plugin.
3. Add observational effects such as the PSF, background noise or saturation with the corresponding `UFig` plugins.
4. Obtain the measured catalog, either by rendering the image, running `SExtractor`, matching the extracted sources to the intrinsic catalog and flagging the sources based on the `SExtractor` output and survey masks, or by using emulators to bypass the image simulation and extraction steps.

5. Save the measured catalog and/or the rendered image.

Apart from the first step, all steps can be done with `UFig` plugins.

## `UFig` images and catalogs

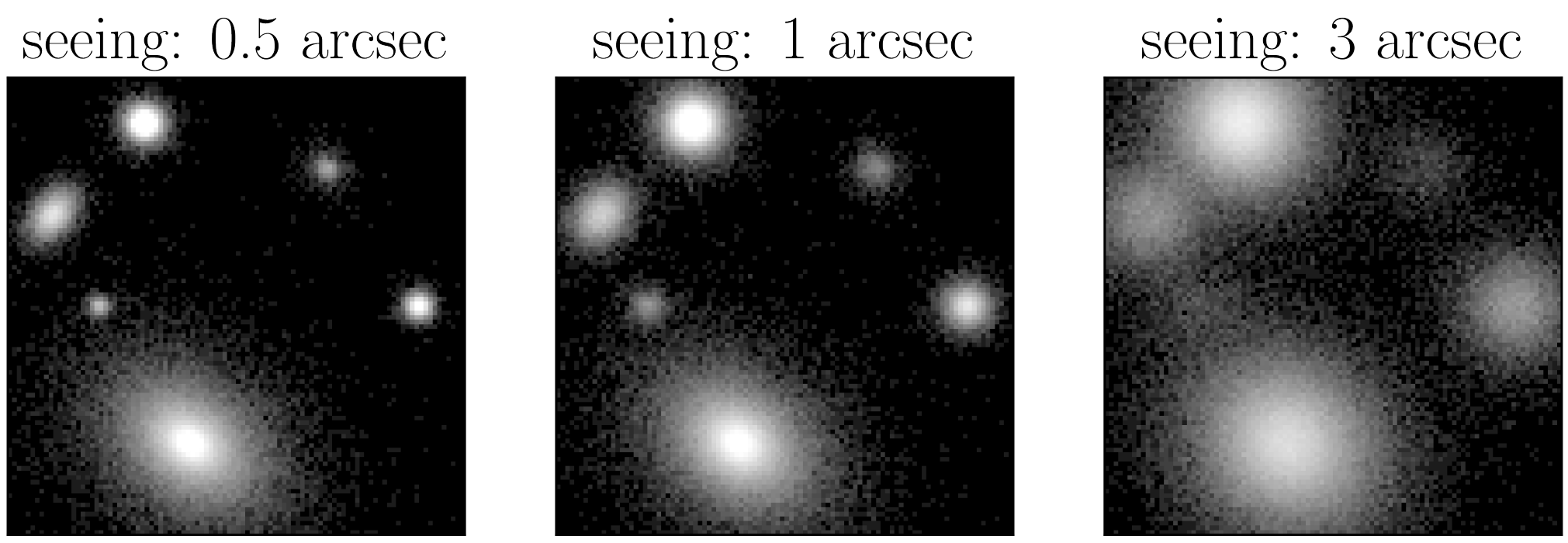

**Figure 1:** Rendered image with three galaxies (a large object at bottom center, an elliptical one at upper-left below a star, and a fuzzier one at upper-right) and three bright stars (round and bright at center-left, center-right, and upper-left). The PSF size varies with different seeing conditions and no background noise is added.

In the simplest case, `UFig` can render an image with a few predefined galaxies and stars without background noise. An example of such a rendered image is shown in Figure 1. From left to right, you see the same objects with different seeing conditions, which change the size of the PSF. The PSF is modelled as a mixture of one or two Moffat profiles $I_i(r)$ given by

$$I_i(r) = I_{0,i} \left(1 + \left(\frac{r}{\alpha_i}\right)^2\right)^{-\beta_i}, \tag{1}$$

with a constant base profile across the image. The ratio of $I_{0,1}$ and $I_{0,2}$ is a free parameter (in the case of a two-component Moffat) and the sum of the two profiles is determined by the number of photons of the object. The $\beta_i$ parameter is free and $\alpha_i$ is chosen such that the half light radius of the profile is one pixel. This base profile is then distorted at each position of an object by three transformations accounting for the size of the PSF, the shape of the PSF (ellipticity, skewness, triangularity and kurtosis) and the position of the PSF (see Herbel et al., 2018 for more details). These distortions can be passed as a constant value across the image, as a map with varying values for each pixel or estimated using the CNN presented in Herbel et al. (2018). Additionally, an approximated brighter-fatter effect can be included by using a first-order linear correction that scales the PSF size and ellipticity components based on the source magnitude relative to a reference magnitude.

Figure 2 shows the same image as in Figure 1 but with added background noise. Background noise can be added as a Gaussian with constant mean and standard deviation across the image or as a map with varying mean and standard deviation for each pixel. Correlated noise is introduced by Lanczos resampling.

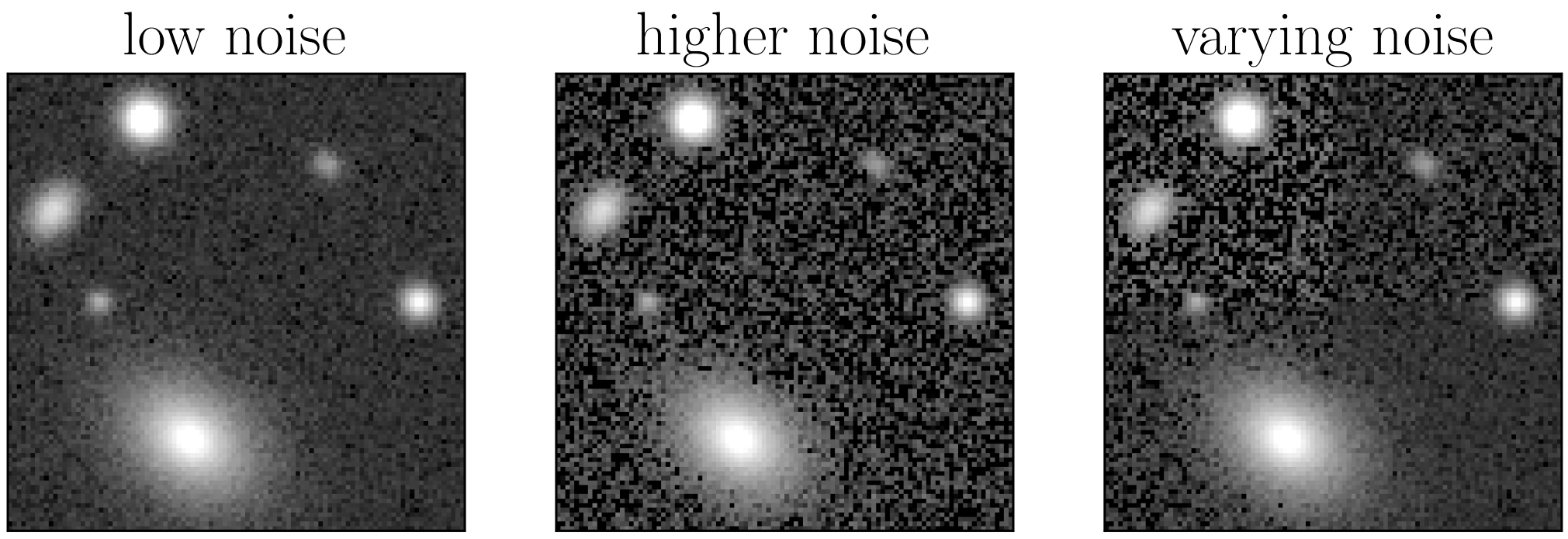

**Figure 2:** Rendered image with three galaxies (a large object at bottom center, an elliptical one at upper-left below a star, and a fuzzier one at upper-right) and three bright stars (round and bright at center-left, center-right, and upper-left). The PSF size is constant and the background level is varied. The left panel shows an image with low background noise, the middle panel higher noise and the right panel shows an image where each quarter has a different background noise.

Creating a more realistic galaxy catalog can be done by using the GalSBI galaxy population model and the corresponding galaxy sampling plugins of the `galsbi` Python package (Fischbacher et al., 2024). An example of rendered images for different bands with galaxies sampled from the GalSBI model presented in Fischbacher et al. (2025) is shown in Figure 3.

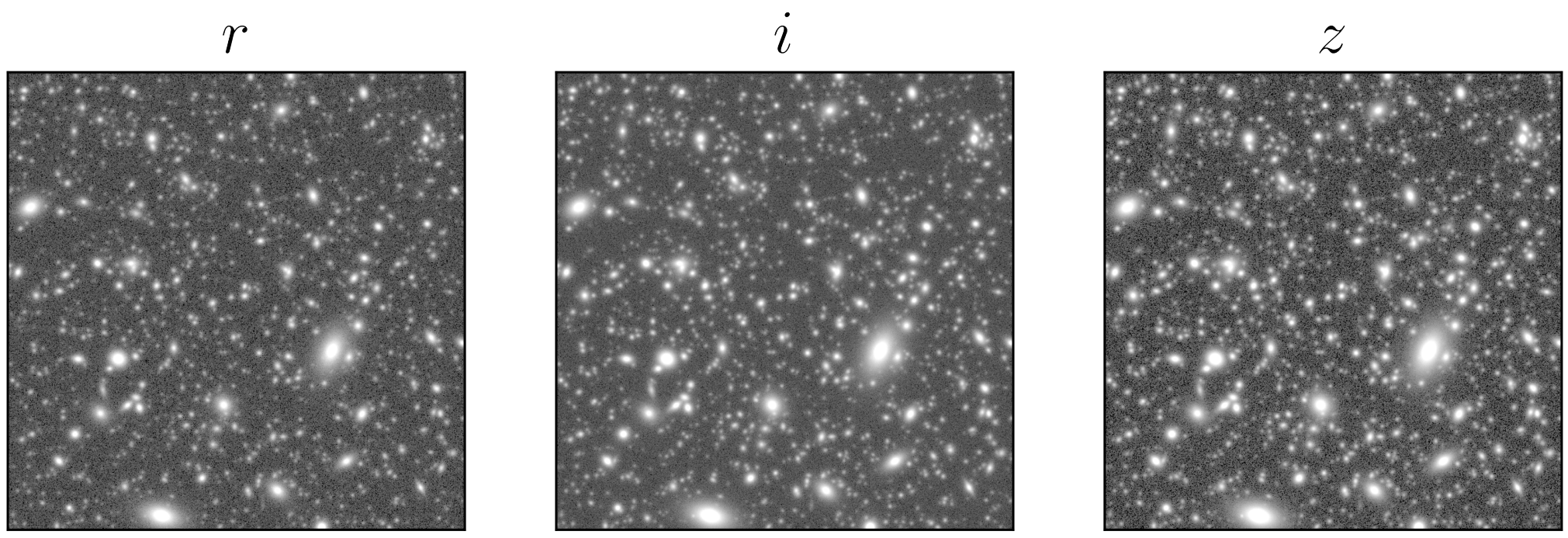

**Figure 3:** Rendered images with galaxies sampled from the GalSBI model for different bands. Background level and PSF estimation correspond to a typical HSC deep field image.

`UFig` also includes plugins to extract sources from the rendered images using `SExtractor` (Bertin & Arnouts, 1996), where the user can specify the `SExtractor` configuration file. This saves the detected objects in a catalog. Figure 4 shows an example of the extracted sources from a rendered image. Stars have a constant size corresponding to the PSF size and ellipticities close to zero as expected for a point source whereas galaxies have broader distributions of sizes and ellipticities.

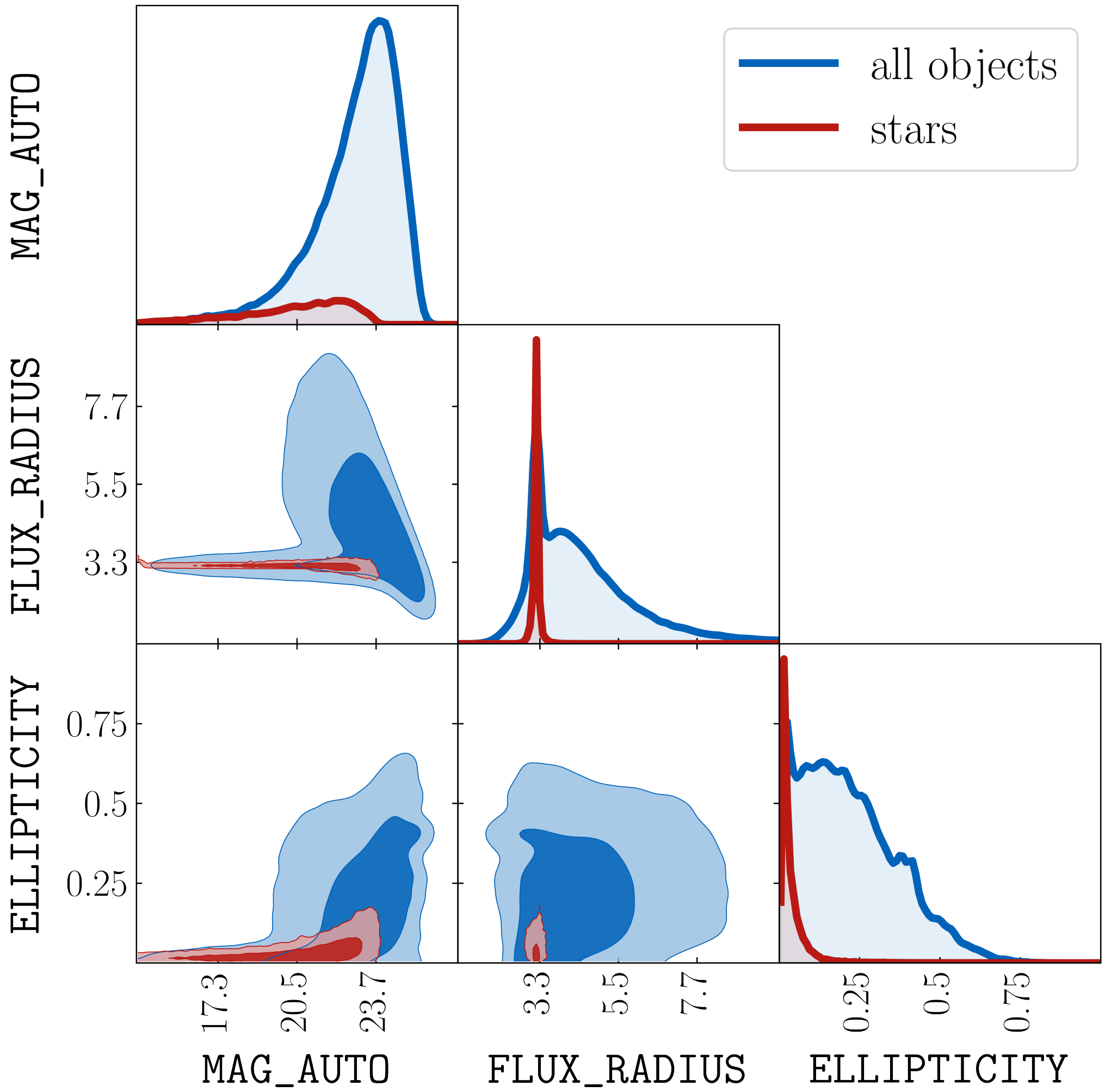

**Figure 4:** Catalog of sources extracted from a rendered image using `SExtractor`. The apparent magnitude (`MAG_AUTO`), angular size in pixel (`FLUX_RADIUS`) and the absolute ellipticity (`ELLIPTICITY`) are shown. All objects in the image are shown in blue, stars are shown in red.

## Acknowledgments

This project was supported in part by grant 200021_143906, 200021_169130 and 200021_192243 from the Swiss National Science Foundation.

We acknowledge the use of the following software packages: NumPy (Van Der Walt et al., 2011), SciPy (Virtanen et al., 2020), Astropy (Astropy Collaboration et al., 2013), `healpy` (Zonca et al., 2019), Numba (Lam et al., 2015), `edelweiss` (Fischbacher et al., 2025), `scikit-learn` (Pedregosa et al., 2011). For the plots in this paper and the documentation, we used Matplotlib (Hunter, 2007), and `trianglechain` (Fischbacher et al., 2023; Kacprzak & Fluri, 2022). The authors with equal contribution are listed in inverse order of their main contribution.